\documentstyle{article}
 
\def\beg{\begin{equation}}
\def\ene{\end{equation}}
\mathchardef\bigtilde="0365
\def\btilde#1{\oalign{$#1$\crcr\hidewidth
    \vbox to.2ex{\hbox{$\bigtilde$}\vss}\hidewidth}}
\def\lsim{\mathrel{\btilde<}}

\begin{document}

\large


$$   $$
\vskip 3cm
\LARGE
\centerline{\bf Modified Laplace transformation method}
\vskip 8pt
\centerline{\bf and its application to the anharmonic oscillator} 
\vskip 1.2 cm
\large
\centerline{\sc Naoki Mizutani$^{1}$ and Hirofumi Yamada$^{2}$}
\vskip 6mm
\centerline{$^{1}$\hskip 3pt\it Research and Development Div. ULVAC JAPAN, Ltd.}
\vskip 4pt 
\centerline{\it 2500 Hagisono, Chigasaki, Kanagawa 253}
\vskip 4pt
\centerline{\it Japan}
\vskip 4pt
\centerline{\it e-mail:qn9n-mztn@asahi-net.or.jp}
\vskip 10pt
\centerline{$^{2}$\hskip 3pt\it Mathematics Department, Chiba Institute of Technology}
\vskip 4pt
\centerline{\it 2-1-1 Shibazono, Narashino, Chiba 275}
\vskip 4pt
\centerline{\it Japan}
\vskip 4pt
\centerline{\it e-mail:yamadah@cc.it-chiba.ac.jp}
\vskip 5mm
\centerline{Feb. 12, 1997}
\vskip 1.5cm
\baselineskip 18pt
\large
\centerline{\bf Abstract}
\vskip 10pt
We apply a recently proposed approximation method to the
evaluation of non-Gaussian integral and anharmonic oscillator.  The method
makes use of the truncated perturbation series by recasting it via the modified 
Laplace integral representation.  The modification of the Laplace transformation is such that the upper limit of integration is cut off and an extra term is added for the compensation.  For the non-Gaussian
integral, we find that the perturbation series can
give accurate result and the obtained approximation converges to the exact
result in the $N\rightarrow \infty$ limit ($N$ denotes the order of
perturbation expansion).  In the case of anharmonic oscillator, we show that
several order result yields good approximation of the ground state
energy over the entire parameter space.  The large order aspect is also
investigated for the anharmonic oscillator.
\newpage
\baselineskip 24pt
\section {Introduction}
\noindent
Anharmonic oscillator is a system that is well understood in both 
perturbative
and non-perturbative aspects$^{1,2}$.  Still, the system plays an important role because it provides a good laboratory for the examination of any calculational scheme newly proposed.  
The Lagrangian of anharmonic oscillator is given by
\beg
L={1 \over 2}\biggl({dq \over dt}\biggl)^{2}-{1 \over 2}m^{2}q^{2}-\lambda
q^{4},
\ene
and the non-linearity is governed by the coupling constant $\lambda$.  The
perturbation expansion is given in powers of $\lambda/m^{3}$.  For instance the ground state energy reads,
\beg
E=m\sum_{n=0}^{\infty}A_{n}\Bigl({\lambda \over m^{3}}\Bigl)^{n},
\ene
where the coefficients are found as$^{1}$
\beg
A_{0}={1 \over 2},\quad A_{1}={3\over 4},\quad A_{2}=-{21 \over 8}\quad 
etc.
\ene
It is known that the coefficient $A_{n}$ grows as $A_{n}\sim
\sqrt{{6 \over \pi^3}}(-3)^{n+1}\Gamma(n+1/2)$ for large $n$ (second paper in ref.1) and thus the series (2) diverges for any
small $\lambda$.  Only for sufficiently small coupling constant,
$\lambda/m^{3}\lsim 0.1$, the series becomes numerically useful by the appropriate truncation. 

The  linear $\delta$ expansion$^{3}$ is one of the framework which leads us to go beyond the weak coupling regime.   For example, recent studies $^{4,5}$ succeeded to approximate the ground state energy in the pure anharmonic case ($m=0$) by using the perturbative series (2).  The strong coupling expansion of
the ground state energy is also given in the literature$^{4}$ (For the convergence issue, see ref.6).  

Recently, a new perturbative scheme was proposed in ref.7.  The
proposed method utilizes the information contained in the perturbation
series
by making Heaviside transformation$^8$ with respect to some parameter (mass in
ref.7).  In the present paper we extend the method and explore how it can be used to obtain non-perturbative result in the two examples, a non-Gaussian integral and the anharmonic
oscillator.   We try to construct approximants for the integral
and the ground state energy.  For example, 
  the approximant of the ground state energy will be given at $N$-th
perturbative
order for $\lambda=1$ as
\beg
E_{approx}(m)=e^{-m^{2}x^{*}}\sum_{n=0}^{N}{A_{n}x^{3n/2-1/2} \over
\Gamma(3n/2+1/2)}+m\sum_{n=0}^{N}{\gamma(3n/2+1/2,m^{2}x^{*}) \over
\Gamma(3n/2+1/2)}{A_{n} \over m^{3n}},
\ene
where $\gamma(p, z)$ denotes the incomplete Gamma function defined by
\beg
\gamma(p, z)=\int^{z}_{0}dt e^{-t}t^{p-1},
\ene
and $x^{*}$ is the "cut-off" parameter to be determined order by order in
some manner.  By our approach one can 
approximate the ground state
energy well over {\it the entire region of the mass} $m$, including the strong coupling limit and the
crossover
region from the weak to the strong coupling regime.  We also note that 
application to
field theories is straightforward.  There exists no renormalization problem.

There is a pioneering work by Graffi et al.$^{9}$, where Borel summation
method was used to compute the energy levels of the
anharmonic oscillator.  Our approach is similar to theirs.  However, our approach is different from it in the respects 
that (i) the variable with respect to which the transformation is carried
out is not the
coupling constant but the mass
square\renewcommand{\thefootnote}{\fnsymbol{footnote}} {\footnote[2]
{\normalsize The mass square, 
$m^{2}$, corresponds to $\lambda^{2/3}$ for the coupling constant.  In
terms of $\lambda^{2/3}/m^{2}$, the original perturbation series is not a
power series.  This is not a negligible difference in our scheme.  The
issue
will be discussed in the last section.}}
and (ii) the Laplace integration is cut off and the extra term is added.  We thus change the representation form itself, while
ref.8 modified the integrand according to Pad$\grave {\rm e}$ construction to make up the Borel sum.  In our case the integrand is just the truncated one.

This paper is organized as follows:  In the next section, we first review
and then extend the method of ref.6.  The extension results in modifying the ordinary Laplace
integral representation as to fit the perturbative approach and enables us to deal with the effect of the explicit mass for both large and small $m^2$.  In section 3
we perform a simple
model calculation of non-Gaussian integral by using the method presented in
the previous section.   We show that the truncated perturbative series re-constructed by
the method gives good approximation of the exact integral and converges to the
exact answer in the $N\rightarrow \infty$ limit ($N$ denotes the order of expansion).  In section 4 we turn to the anharmonic oscillator.  Via
perturbation series, the ground
state energy is approximately calculated 
for various $\lambda/m^{3}$ to several higher orders.  We will show that,
already at 5-th order, the error of our approximation is less than 1 percent for any $\lambda/m^{3}$.  In section 5 we
address to large order aspects of our approach by proceeding to 249-th order. 
Some discussion and summary of the present work is given in section 5.  In appendix we state the relation
between our method and the linear $\delta$ expansion in a particular limit.

\section{Modified Laplace transformation}
\noindent
A part of the content of this section is same as the corresponding part
in ref. 6.  However, to make the presentation self-contained, we 
allow some overlaps with the work. 
 
For a given physical function $f(\sigma)$, we consider the Heaviside
transform$^{7}$ given by the Bromwich
integral,
\begin{equation}
\hat f(x)=\int^{p+i\infty}_{p-i\infty}{d\sigma \over 2\pi i}
\exp(\sigma x){1 \over \sigma}f(\sigma).
\end{equation}
Here the parameter $p$ represents the location of the vertical contour.  
Although $m^{2}$ corresponds to $\sigma$ in the cases we work with, there
would be other choices in general.  
The contour of integration should be placed on the right of all the possible poles and the
cut of $f(\sigma)/\sigma$.  Then, if $x<0$, the contour may be
closed into the right half circle and $\hat f(x)$ is found to vanish.  From $\hat f(x)$ we have $f(\sigma)$ via the
Laplace integral of the second kind,
\begin{equation}
f(\sigma)=\sigma\int^{\infty}_{-\infty}dx\exp(-\sigma x)\hat f(x).
\end{equation}
Since $\hat f(x)=0$ when $x<0$, the integration range reduces to
$[0,\infty)$. 
However it is convenient to keep the range as $(-\infty,+\infty)$ to handle
partial integration easily.  

Now suppose that one is interested in the value of $f$ at $\sigma=0$, but the perturbative expansion of $f(\sigma)$ does not allow one to let $\sigma$ arbitrary small.    As in
the Fourier transformation, small $\sigma$ behavior of $f(\sigma)$ is
connected with the large $x$ behavior of $\hat f(x)$.  More precisely we
find that
\beg
\lim_{\sigma\rightarrow +0}f(\sigma)=\lim_{x\rightarrow \infty}\hat f(x),
\ene
where the existence of both limits are assumed.  Our
approximation procedure is based on (8) and it goes as follows.  
 Let $f_{N}(\sigma)$ denotes the perturbative
expansion of $f(\sigma)$ to $N$-th order.  Then the corresponding Heaviside
function is given by
\beg
\hat f_{N}(x)=\int^{p+i\infty}_{p-i\infty}{d\sigma \over 2\pi i}
{\exp(\sigma x) \over \sigma}f_{N}(\sigma).
\ene 
In cases of our interest, we can not take the naive limits, $\sigma\rightarrow 0$ or $x\rightarrow \infty$, in the both functions.  Then there would be two routes to approximate $f(0)$.
  Naive one is to approximate it by fixing $\sigma$ as small
as possible
in some manner.  The other is, relying upon (8), to approximate $f(0)$ by
$\hat f_{N}(x)$ where $x$ should be fixed at some large value, $x^{*}$. 
Here note that 
$\hat f(x)$ often has larger convergence radius than $f(\sigma)$.  For example,  if $f(\sigma)=\sum_{n=0}^{\infty}a_{n}/\sigma^{n}$, then $\hat
f(x)=\sum_{n=0}^{\infty}a_{n}x^{n}/n!$ ($x>0$).  
When $\hat f(x)$ has the convergence radius larger 
than that of $f(\sigma)$, it would be convenient to deal with $\hat f_{N}$ rather than $f_{N}$.  This is because 
we can probe the
large $x$ behavior of $\hat f$ by $\hat f_{N}$ so that we have more chance to know the accurate value of
$\hat f(\infty)$ and $f(0)$ accordingly.  Therefore, we choose to 
approximate $f(0)$ by $\hat f_{N}(x^{*})$.
   The explicit way of fixing $x^{*}$ is discussed in the next section. 
Here
we just mention that, from the estimation of the upper bound of reliable 
perturbative region, $x^{*}$
will be fixed by the stationarity condition,
\beg
{\partial \hat f_{N}(x) \over \partial x}\biggl|_{x=x^{*}}=0.
\ene
If there are several solutions, we should input the largest $x^{*}$ into
$\hat f_{N}$.  This is 
obvious because the value of $\hat f$ at $x=\infty$ is what we are looking for.

The above approach can be extended to the approximation of the function
itself
over the entire region of $\sigma$.  Let us start the discussion by showing
how we can approximate the small $\sigma$ expansion of $f(\sigma)$,
\beg
f(\sigma)=f(0)+f^{(1)}(0)\sigma+f^{(2)}(0){\sigma^{2} \over 2!}+\cdots.
\ene
As well as $f(0)$, we can approximate $f^{(k)}(0)$ as the following manner: 
From the formulas,
\begin{eqnarray}
\sigma{\partial f(\sigma) \over \partial \sigma}&\rightarrow &
-x{\partial \hat f(x) \over \partial x}\nonumber\\
{1 \over \sigma}f(\sigma)&\rightarrow &\int^{x}_{-\infty}dy\hat f(y),
\end{eqnarray}
where the rightarrow represents the Heaviside transformation, 
we have
\beg
f^{(k)}(\sigma)\rightarrow \int^{x}_{-\infty}dy(-y)^{k}{\partial \hat f(y)
\over \partial y}\stackrel{\rm def}{=}\alpha_{k}(x).
\ene
Assuming the expansion (11), the above two functions agree with each other
at $\sigma=0$ and
$x=\infty$.  Since what we have at hand is the perturbative one,
$f_{N}^{(k)}(\sigma)$, the derivative $\alpha_{k}(x)$ is also truncated at
order $N$.  Hence, as in the previous case, we approximate $f^{(k)}(0)$ by
$\alpha_{k}(x)$ by
fixing the upper limit of integration $x$ as large as possible within the
perturbative region.  Replacing $\hat f(y)$ by $\hat f_{N}(y)$, we
then choose the input $x$, say $x^{*}_{k}$, according to
the same logic as that for $x^{*}$.  That is, $x^{*}_{k}$ is determined by
the stationarity condition,
\beg
{\partial \alpha_{k} \over \partial x_{k}^{*}}={\partial \over \partial
x_{k}^{*}}
\int^{x^{*}_{k}}_{-\infty}dy(-y)^{n}{\partial \hat f_{N}(y) \over \partial
y}=
(-x_{k}^{*})^{k}{\partial \hat f_{N}(x_{k}^{*}) \over \partial x_{k}^{*}}=0.
\ene
We find that for any $k$
\beg
x_{k}^{*}=x^{*}
\ene
where $x^{*}$ is a solution of (10).  Substituting $x^{*}$ into $\alpha^{(k)}(x)$ we can construct the approximate Taylor expansion, 
\beg
f(\sigma)\sim \sum_{k=0}^{\infty}{\alpha_{k}(x^{*}) \over
k!}\sigma^{k}=\sum^{\infty}_{k=0}{\sigma^{k} \over k!}
\int^{x^{*}}_{-\infty}dx(-x)^{k}{\partial \hat f_{N}(x) \over \partial x},
\ene
where $\alpha_{0}(x^{*})=\hat f_{N}(x^{*})$.  

Now it is an easy
task to obtain expression which can be used for entire $\sigma$ region: 
We observe that the right hand side of (16) is easily summed to $\int^{x^{*}}_{-\infty}dx e^{-\sigma
x}\partial \hat f / \partial x$.  Then, 
by integrating by parts using $\hat f(x)=0$ for $x<0$, it is written as
\beg
e^{-\sigma x^{*}}\hat
f_{N}(x^{*})+\sigma\int_{-\infty}^{x^{*}}dx
e^{-\sigma x}\hat f_{N}(x)\stackrel{\rm def}{=}f_{N}(\sigma, x^{*}).
\ene
The left hand side of (17) defines the modification of Laplace
transformation.  
Note that, the naive $x^{*}\rightarrow
\infty$ limit recovers the ordinary Laplace
transform and gives $f_{N}(\sigma)$.  We point out, however, that for small $m$ the dominant
contribution comes from the first term.  Actually the second term, the
cut-off Laplace integral, gives zero in the $m\rightarrow 0$ limit.  Thus
the first term is the crucial ingredient in our approach.  To summarize, our
approach results in approximating $f(\sigma)$ by
$f_{N}(\sigma,x^{*})$ defined by (17). 

It would be interesting to see how the approximant, $
f_{N}(\sigma,x^{*})$, is different from the ordinary perturbative series. 
Let us rewrite the integral in (17) as
\beg
\sigma\int_{-\infty}^{x^{*}}dx
e^{-\sigma x}\hat f_{N}(x)= f_{N}(\sigma)-\sigma\int_{x^{*}}^{\infty}dx
e^{-\sigma x}\hat
f_{N}(x).
\ene
Then we have
\beg
 f_{N}(\sigma,x^{*})=f_{N}(\sigma)+f_{N}^{corr}(\sigma,x^{*}),
\ene
where
\beg
f_{N}^{corr}(\sigma,x^{*})=e^{-\sigma x^{*}}\hat
f_{N}(x^{*})-\sigma\int_{x^{*}}^{\infty}dx e^{-\sigma x}\hat f_{N}(x).
\ene
If $f_{N}(\sigma)$ is given as $f_{N}=\sum^{N}_{n=0}a_{n}/\sigma^{n}$, then
$f_{N}^{corr}$ is given by
\beg
f_{N}^{corr}=e^{-\sigma x^{*}}\sum^{N}_{n=0}{a_{n}x^{*n} \over
n!}-\sum^{N}_{n=0}{a_{n}\Gamma(n+1,\sigma x^{*}) \over n! \sigma^{n}},
\ene
where
\beg
\Gamma(z,p)=\int^{\infty}_{p}dt e^{-t}t^{z-1}.
\ene
Using the asymptotic expansion of $\Gamma(z,p)$, 
\beg
\Gamma(z,p)=p^{z-1}e^{-p}\biggl[1+\sum^{\infty}_{k=1}
{1 \over p^{k}}(z-1)(z-2)\cdots(z-k)\biggl],
\ene
we find
\beg
f_{N}^{corr}=-e^{-\sigma x^{*}}\sum_{k=1}^{\infty}{1 \over \sigma
^{k}}{\partial^{k} \hat f_{N}(x^{*}) \over \partial x^{*k}}.
\ene
From its structure, (24) gives the large $\sigma$ expansion of
$f^{corr}_{N}$. 
We find that under the stationarity
condition (10) the first term in (24) vanishes.  This shows
that, for large $\sigma$, the condition minimizes the deviation of the
approximant from the ordinary
perturbative result. 

\vskip 10pt
\section{A simple model calculation: non-Gaussian integral}
\noindent 
It would be worthwhile carrying out the calculation for a solvable model to
gain some concrete feeling on the method to be used.  We take up here a
non-Gaussian integral,
\beg
Z(m, \lambda)=\int^{\infty}_{-\infty} dq e^{-m^{2}q^{2}-\lambda q^{4}}.
\ene
The perturbation series is given
by expanding the integral in terms of $\lambda$.  The result reads
\beg
Z=\sum^{\infty}_{n=0}{(-\lambda)^{n} \over n!}{\Gamma(2n+1/2) \over
m^{4n+1}}.
\ene
It is easy to see that the series diverges for any $\lambda/m^{4}$.  Only
when $\lambda/m^{4}\lsim 0.1$, the series can be put into numerical use by the appropriate truncation.   
Nevertheless, we will show that we can obtain the strong coupling
($\lambda/m^{4}\gg1$) expansion from the divergent weak coupling series, and
even an approximant which is effective over
the entire coupling regime.

First step is to obtain the integrand of Laplace representation.   This is done by calculating the integration (6) over $m^{2}$
(i,e., $\sigma=m^{2}$).  Then the result is given by
\beg
\hat Z(x,\lambda)=\sum^{\infty}_{n=0}{(-\lambda)^{n} \over n!}{x^{2n+1/2}
\over (2n+1/2)}\theta(x)=\lambda^{-1/4}\sum^{\infty}_{n=0}{(-1)^{n} \over
n!}{(\sqrt{\lambda}x)^{2n+1/2}
\over (2n+1/2)}\theta(x),
\ene
where $\theta(x)$ denotes the ordinary step function ($\theta(x)=1$ for
$x>0$ and $0$ for $x<0$).  Note that $\hat Z$
converges for any $\lambda x^{2}$.  For the sake of notational simplicity we set $\lambda=1$ hereafter.

Let us introduce the perturbatively truncated $\hat Z$ by
\beg
\hat Z_{N}(x,\lambda=1)=\hat Z_{N}(x)= \sum^{N}_{n=0}{(-1)^{n} \over
n!}{x^{2n+1/2} \over
(2n+1/2)}\theta(x).
\ene
Now, we turn to the approximation of $Z(m, \lambda=1)=Z(m)$ by using (28). 
We first discuss how to choose the input $x^{*}$.  The break down of
perturbative series would generally appear as the rapid rise or fall of the Heaviside function due to the domination of the highest term in (28). 
Actually, the graphs of the function $\hat Z_{N}(x)$ to
the
first several orders show it is the case (see Fig.1).  Then, since $\hat
Z_{N}(x)$ is an
alternative series, the rise and fall occurs alternatively.  For example,
for odd
$N$, $\hat Z_{N}(x)$ temporally increases with $x$ in the reliable perturbative
region, but after that (eventually?) it falls down to $-\infty$.  We note that $\hat Z_{N}$
behaves temporarily flat just before the break down as shown in Fig.1.  In this
case the limit of perturbation would emerge typically as the stationary
point.  This is the reason we employ the stationarity condition,
\beg
0={\partial \hat Z_{N}(x) \over \partial x}=\sum^{N}_{n=0}{(-1)^{n} \over
n!}x^{2n-1/2}\theta(x)+\sum^{N}_{n=0}{(-1)^{n} \over n!}{x^{2n+1/2} \over
(2n+1/2)}\delta(x),
\ene
to fix the perturbative limit, $x^{*}$ ($\delta(x)$ denotes the Dirac
delta function).  Dropping $\theta(x)$ and $\delta(x)$ which are
irrelevant, we have
\beg
\sum^{N}_{n=0}{(-1)^{n} \over n!}x^{2n-1/2}=0.
\ene
The solution of (30) exists for odd $N$ and depends on $N$.  By substituting $x^{*}$ into $\hat Z_{N}(x)$ we have the approximant of $Z(0)$.  The obtained result is satisfactory
as shown in Table 1.   Note that $x^{*}$ increases with $N$, which is a desirable result.   

 We can also approximate the small mass (strong coupling) expansion of $Z(m)$: 
First
note that the coefficients of the $k$-th power of $m^{2}$, $\alpha_{k}(x^{*})$, is given by 
\beg
\alpha_{k}(x^{*})=\int^{x^{*}}_{-\infty}dx (-x)^{k}{\partial \hat Z_{N}(x) \over
\partial
x}=(-1)^{k}\sum^{N}_{n=0}{(-1)^{n}(x^{*})^{k+2n+1/2} \over n! (k+2n+1/2)}.
\ene
Thus, at 15-th order for example, we have
\begin{eqnarray}
Z_{N}(m,x^{*})&=&\sum^{\infty}_{k=0}{\alpha_{k} \over k!}m^{2n}\nonumber\\
 &=&1.811655
-0.609988 m^{2}+
0.223363 m^{4}
-0.074001 m^{6}+
0.022042 m^{8}+\cdots.
\end{eqnarray}
The exact result reads from (25) that $Z(m)={1 \over
2}\sum_{n=0}^{\infty}\Gamma(n/2+1/4)(-m^{2})^n/n!$ and is given numerically
as 
\beg
Z(m)=1.812805
-0.612708 m^{2}+
0.226601 m^{4}
-0.076589 m^{6}+
0.023604 m^{8}+\cdots.
\ene
We see the good agreement of (32) and (33) up to several orders.  Thus,
we reach sufficient accuracy of strong coupling expansion from only the
information of the truncated weak coupling series.

Next we examine the approximation of $Z(m)$ for various $m$.   The approximant is given from (28) and (17) as,
\beg
Z_{N}(m, x^{*})=e^{-m^{2}x^{*2}}\sum^{N}_{n=0}{(-1)^{n} \over
n!}{(x^{*})^{2n+1/2} \over (2n+1/2)}+\sum^{N}_{n=0}{(-1)^{n} \over
n!(2n+1/2)}{\gamma(2n+1/2,m^{2}x^{*2}) \over m^{4n+1}}.
\ene     
Table 2 shows the
value of the approximant at $N=15$ together with
the exact results.  We find that
the approximant gives good values for the sample of $m$.

Before closing this section we show that the approximant $\hat Z_{N}(x^{*})$ converges to $\hat Z(\infty)$ in the $N\rightarrow \infty$ limit.  First we show the convergence of
$\lim_{x\rightarrow \infty}\hat Z(x)$, where
$\hat Z(x)$ denotes the exact transformed function, by calculating Bromwich integral exactly.  From (6) and (25) $\hat Z(x)$ is given as
\beg
\hat Z(x)=2\int^{\sqrt{x}}_{0}dq e^{-q^{4}}={1 \over 2}\gamma(1/4,x^{2}),
\ene
where we used that $e^{-m^{2}q^{2}}$ transforms to $\theta(x-q^{2})$.  Thus it is apparent
that $\lim_{x\to \infty}\hat Z(x)=\Gamma(1/4)/2$,
which agrees with $Z(0)$.  

For the proof we need to know how $x^{*}$ behaves for large $N$.  The
relation is found as follows:  We note that the condition (30)
is viewed as the truncation of the equation, $x^{-1/2}\exp(-x^2)=0$.  Since the
series expansion of $\exp(-x)$ has infinitely large convergence radius,
the obtained solution tends to $+\infty$ as $N\rightarrow \infty$.  More precisely, by
assuming the form, $x^{*2}\sim aN^{b}$ ($a,b$: constant), we find the following
scaling at large $N$,
\beg
x^{* 2}\sim {1 \over 3}N^{1}.
\ene

Now, let us define the reminder, $\hat R_{N}$,
by
\beg
\hat R_{N}(x)=\hat Z_{\infty}(x)-\hat Z_{N}(x)=\sum^{\infty}_{n=N+1}{(-1)^{n} \over
n!}{x^{2n+1/2} \over
(2n+1/2)}.
\ene
Since $\hat Z_{\infty}(x)$, the perturbative series
to all orders, apparently converges to exact $\hat Z(\infty)$ because of the infinite convergence radius, it is
sufficient to show that 
\beg
\lim_{N\rightarrow \infty}\hat R_{N}(x^{*})=0.
\ene
This is easily verified:  
Using the Stiring's formula, we obtain
\beg
|\hat R_{N}(x^{*})|<\sum^{\infty}_{n=N+1}{e^n(x^{* 2})^{n+1/4} \over 2\sqrt{2
\pi}N^{n+3/2}},
\ene
and from (36) we then find
\beg
|\hat R_{N}(x^{*})|<{(e/3)^{5/4} \over \sqrt{8\pi} (1-e/3)}N^{-5/4}\Bigl({e \over
3}\Bigl )^{N}\rightarrow 0\quad
(N\rightarrow \infty),
\ene
which proves (38).

\section{Approximation of the ground state energy of the anharmonic
oscillator}
\noindent 
We turn to discuss the anharmonic oscillator from this section.  In
complicated systems it is generally hard to proceed to arbitrary higher
orders.  Hence it is practically important to study whether an employed
approximation scheme works at low orders.

In this section we use perturbative series up to 9-th order and show that
the
several low order result can yield good approximation of the ground state
energy via modified Laplace representation.  

Let $\lambda=1$ for notational simplicity.  Our first task is to obtain the
Heaviside function of the perturbative ground state energy, $E_{N}(m)$,
\beg
E_{N}(m)=\sum_{n=0}^{N}{A_{n} \over (m^{2})^{3n/2-1/2}}.
\ene
Heaviside transform of 
$m^{-3n+1}$ with respect to $m^{2}$ gives
\beg
m^{-3n+1}\rightarrow {x^{3n/2-1/2} \over \Gamma(3n/2+1/2)}\theta(x).
\ene
Then from (41) and (42) we have
the Heaviside function, $\hat E_{N}(x)$,
\beg
\hat E_{N}(x)=\sum_{n=0}^{N}{A_{n}x^{3n/2-1/2} \over \Gamma(3n/2+1/2)}\theta(x).
\ene
For notational simplicity we omit the step function in what follows. 

 By suitable replacement of variable, the
function
$\hat E$ agrees with the function
appeared in ref. 5 in which the $\delta$ expansion method was applied to
the anharmonic oscillator (see eq.(7) of ref. 5).  Authors of ref. 5
considered the case where the 
coupling, $\delta \Omega^{2}$, grows to large orders ($\delta$ represents a fictitious parameter
which is to be set 1 at the end of calculation and $\Omega^{2}$ denotes the
mass auxiliary
introduced to divide the given Lagrangian un-conventionally).  Then the
authors found that their approximant converges
to the form (43) in that large order limit\renewcommand{\thefootnote}{\fnsymbol{footnote}} {\footnote[3]
{\normalsize  In $\delta$ expansion scheme, this limit corresponds to
infinite perturbative order.  Then we note that in this limit the parameter
$\Omega$ is no longer arbitrary because it should be of the same order of
magnitude with the perturbative order.  We also point out that this limit
was considered in ref.9 from different motivation. }}.  This implies that
there may be some connection between the
linear $\delta$ expansion and our method.  Actually we show in appendix that it is the case;  We will show that, in a particular limit, the Heaviside transform is induced in a suitable interpretation. 

Now, let us discuss the energy approximation using (43).  As in the previous section we use the following condition,
\beg
{\partial \hat E_{N}(x) \over \partial x}=0,
\ene
to fix the input $x^{*}$.  
Up to $N=9$, while there is no solution for even $N$, we find just one
solution for each odd $N$.  By substituting $x^{*}$ into $\hat E_{N}(x)$, we obtain the
approximation of $E(0)$ which is the energy in the pure anharmonic case or in other words in the strong coupling limit.  We also evaluate the
succeeding Taylor coefficients of first five terms.  For the purpose it may be convenient to use the integrated form of $\alpha_{k}$,
\beg
E^{(k)}(0)\sim \alpha_{k}=(-1)^{k}\sum^{N}_{n=0}{A_{n}(3n/2-1/2)(x^{*})^{k+3n/2-1/2}
\over \Gamma(3n/2+1/2)(k+3n/2-1/2)}\hskip 5mm (k=1,2,3,\cdots),
\ene
which comes from substitution of (43) into the general formula (13).  
Then, we have the strong coupling expansion,
\beg
E(m)\sim \alpha_{0}+{\alpha_{1} \over 1!}m^{2}+{\alpha_{2} \over
2!}m^{4}+\cdots,
\ene
where $\alpha_{0}=\hat E_{N}(x^{*})$.  We have done numerical calculation
with Mathematica.   
As shown in Table 3, the obtained results agree well with the recent  result reported by Kleinert$^{4}$.  
Next we check the approximation for various $m$.  For
the purpose we substitute (43) into (17) and use the definition of
incomplete Gamma
function (5) to obtain the approximant, $E_{N}(m,x^*)$.  The result reads at the order $N$ as
\beg
E_{N}(m,x^{*})=e^{-m^{2}x^{*}}\sum_{n=0}^{N}{A_{n}x^{3n/2-1/2} \over
\Gamma(3n/2+1/2)}+m\sum_{n=0}^{N}{\gamma(3n/2+1/2,m^{2}x^{*}) \over
\Gamma(3n/2+1/2)}{A_{n} \over m^{3n}}.
\ene
We find that the numerical calculation shows the good results for all $m$ \renewcommand{\thefootnote}{\fnsymbol{footnote}}
{\footnote[4]
{\normalsize The reference value is generated according to the method
of ref.11.  We thank H. Suzuki for the correspondence of the reference.}}.  From these results, we find that the accuracy of calculated
energy is quite satisfactory.  Already at the 1-st order, the approximation
gives error only within 11 percent for all $m$.  And at 5-th order, the
error is less than 1 percent.  These results are depicted in Fig.2.  In
particular
we note that $E_{N}(m, x^{*})$ improves the approximation of $E$ in the
crossover region
of weak and strong coupling regimes, $0.1\lsim 1/m^{2}\lsim 1$.  Note that in this
intermediate coupling region the ordinary
perturbation series can not be used because the series never be close to the exact value under any truncation.  

It would be interesting to see the explicit difference of our approximant
from the ordinary
perturbative series. Repeating the steps from (18) to (24), we find
\beg
E_{N}(m,x^{*})=E_{N}(m)+E_{N}^{corr}(m,x^{*}),
\ene
where
\beg
E_{N}^{corr}(m,x^{*})= e^{-m^{2}x^{*}}\hat
E_{N}(x^{*})-m\sum_{n=0}^{N}{\Gamma(3n/2+1/2,m^{2}x^{*}) \over
\Gamma(3n/2+1/2)}{A_{n} \over
m^{3n}}=-e^{-m^{2}x^{*}}\sum^{\infty}_{i=1}{b_{i,N}
\over
(m^{2})^{i}},
\ene
and
\beg
b_{i,N}=\sum_{n=0}^{N} {A_{n}(x^{*})^{3n/2-1/2-i} \over
\Gamma(3n/2+1/2)}\prod_{j=1}^{i}(3n/2+1/2-j)={\partial^{i}\hat E_{N}(x^{*})
\over
\partial x^{*i}}.
\ene
The 
coefficients slowly changes order by order except for $b_{1,N}$ which
satisfies from (44), 
\beg
b_{1,N}=0.
\ene
We have done numerical computation of $b_{i,N}$ at $N=1,3,5,7,9$.  To
several $i$, we find that the size of $b_{i,N}$ decreases as $N$ increases.  This is an expected tendency.  
In fact, we can show that $b_{i,N}$ should converges to zero in the
$N\rightarrow\infty$ limit.  The basic relation to be noted is that
\beg
\sigma f(\sigma) \rightarrow {\partial \hat f(x) \over \partial x}.
\ene
From (52), it is easy to see that
\beg
\sigma^{i}f(\sigma)\rightarrow {\partial^{i} \hat f(x) \over \partial
x^{i}},
\ene
and for our case,
\beg
m^{2 i}E(m)\rightarrow {\partial^{i} \hat E(x) \over \partial x^{i}}.
\ene
There should be the agreement condition (8) between these functions and
therefore, noting that $\lim_{m^{2}\rightarrow 0}m^{2 i}E(m)=0$, the
coefficients $b_{i,N}$ should tend to zero as $N$ increases to infinity if
the approximation procedure is working well.  This issue will be studied further in the next section.

\section{Higher order behavior}
\noindent
The important information in our approach is contained in the Heaviside
function $\hat E(x)$.  We therefore investigate the properties of $\hat E(x)$ relevant
to our analysis by extending the perturbative order up to 249-th.

An important issue in our
approach is whether $\lim_{x\rightarrow \infty}\hat E(x)$
exists or not.  Although we do not have rigorous proof, we see
convincing answer by figuring out $\hat E_{N}(x)$ to large $N$.  We have
generated perturbative coefficients $A_{n}$ up
to 249 terms with the help of Mathematica and plotted the graph of $\hat
E_{249}(x)$ as shown in Fig.3.   We note that
plateau starts around $x\sim 1$ and abruptly grows up around $x\sim 3.2$, which shows the break down of perturbation expansion.  Taking closer look, we find that the function $\hat E_{249}(x)$
weakly oscillates at the plateau region.    
The amplitude of the oscillation is very tiny indeed; The difference between the first extremum
and the next is just $0.0000103\cdots$ which should be compared
with the first extremum value, $0.667975902279\cdots$.  The difference between
the second and third is about $0.00000001$.  Thus the amplitude decreases as the function oscillates to larger $x$.  The values of three stationary points of $\hat E_{249}$ are given as 
\begin{eqnarray}
\hat E_{249}&=&0.667975902279\cdots, \hskip 5mm x=1.139689002700\nonumber\\
\hat E_{249}&=&0.667986268727\cdots, \hskip 5mm x=2.069065340532\nonumber\\
\hat E_{249}&=&0.667986259143\cdots, \hskip 5mm x=2.987637042160.
\end{eqnarray}
These values shows how 
the $\hat E_{N}$ at the
plateau is close to the value $E(0)$ which is known to be$^{12}$
\beg
E(0)=0.667986259155777108270962\cdots.
\ene
Thus the behavior of $\hat E_{249}(x)$ for $x\le 3$, where the function is reliable, strongly suggests that $\lim_{x\to \infty}\hat E(x)$ would exist and consequently agree with $E(0)$.

The first, second and third stationary points begins to appear from 28,101 and 246 orders, respectively.  Hence, several solutions exist for some orders higher than $28$-th.  This phenomenon was also observed in the $\delta$ 
expansion framework$^{5}$.  There it was found that the value of interest at
largest $1/\Omega$ was most accurate.  But within the framework there is no a priori reason why one should take the largest $1/\Omega$.  On the other hand, it
is obvious in our approach that one should focus on the largest
$x^{*}$ as we mentioned in section 2.  The behavior of our approximants $\hat E_{N}(x^{*})$ as the order increases is as follows.  For $x$ smaller than the largest $x^{*}$, $\hat E_{N}(x)$ is a good approximation of the exact function and the departure starts around $x\sim x^{*}$.  Therefore the largest stationary point sits in the vicinity of the exact function, and slides to larger $x$ direction along the curve of the exact function as $N$ increases.  Hence the convergence issue of the approximants, $\{\hat E_{N}(x^{*}) | N=1,2,\cdots\}$, is tightly connected with the convergence of $\lim_{x\to \infty}\hat E(x)$ and how the stationary solution, $x^{*}$, grows with the order.  Since, by the definition, the largest $x^{*}$ is located at the upper limit of reliable region, $\{\hat E_{N}(x^{*})| N=1,2,\cdots\}$ would converge to $\hat E(\infty)$.  It is also clear that $\hat E_{N}(x^{*})$ oscillates as $N$ increases by  following the function $\hat E(x)$.  This oscillatory property of the approximant was observed (but not clarified) by Kleinert$^{4}$. 

As is obvious from the above discussion, the largest stationary point around the order $N=246$ (at which the third stationary point of $\hat E(x)$ is settled) is approximately given by the third stationary point shown in the last of (55).  Therefore using that value of $x^{*}$, we can see how accurate $E_{N}(m,x^{*})$ is for various $m^{2}$.  The result of computer calculation is shown in Table 4 and shows that the obtained values are quite accurate for all $m^{2}$ .

Finally let us comment on the behavior of $b_{i,N}$, the coefficients of
$E_{N}^{corr}$ at large $m$, for large $N$.  We have calculated them at 
$N=28,101,249$.  We find that the size of
$b_{i,N}$ decreases to zero as the order $N$ increases.   For example, the
results for $i=1$ to $7$ at $N=249$ are respectively given as
follows;
\begin{eqnarray}
b_{1,249}&=&0{\rm E}{-22},\hskip 4pt b_{2,249}=
8.259931{\rm E}{-10},\hskip 4pt b_{3,249}=
-8.27746{\rm E}{-9},\hskip 4pt 
b_{4,249}=5.094257{\rm E}{-7},\nonumber\\
b_{5,249}&=&
4.804239{\rm E}{-5},\hskip 4pt b_{6,249}=
6.054357{\rm E}{-3},\hskip 4pt b_{7,249}=
7.451039{\rm E}{-1},
\end{eqnarray}
Thus, for lower $i$, the result almost agrees with the requirement that
$\lim_{x\rightarrow
\infty}\partial^{i} \hat E(x)/\partial x^{i}=0$. 
  For larger $i$, however, the coefficient $b_{i,249}$ grows rapidly.  This represents that as $x$ exceeds $x^{*}$ $\hat E_{249}(x)$ abruptly grows as shown in Fig.3.

\section {Discussion and conclusion}
\noindent 
Critical things leading our scheme to the success are that the convergence radius, $\rho$, is infinite for the Heaviside functions $\hat f(x)$ and that they
quickly approach to the value $\hat f(\infty)$ at finite $x$.  
The later fact is confirmed numerically for the anharmonic oscillator from the remarkable closeness of $\hat E_{N}(x)$ at
$x\in[1,3]$ to $\hat E(\infty)$.  For the non-Gaussian integral it is  
analytically clarified as follows:  By using
$\gamma(1/4,x^{2})=\Gamma(1/4)-\Gamma(1/4,x^{2})$ and
the expansion (23) we obtain from (35) that
\beg
\hat Z(x)={1 \over 2}\Biggl[ \Gamma(1/4)-x^{-3/2}\exp(-x^{2})
\biggl(1+\sum_{i=1}^{\infty}x^{-i}\prod_{j=1}^{i}(1/4-i)\biggl)\Biggl].
\ene
The approach to $\hat f(\infty)$ is speedy because
of exponential damping at large $x$.  It is interesting to note that $\hat
Z(x)$ has no power-like term, $x^{-n}$, in large $x$ expansion (58).

Actually these features are related with the choice of integration variable generally denoted as $\sigma$. 
We discuss on this issue by our two examples.  

First consider the non-Gaussian integral case and let $\sigma=m^{\beta}$.  We numerically study 
on how the function $\hat Z$ varies according to the power $\beta$. 
In calculating the transform, the following result is convenient to use:
\beg
\sigma^{\xi}\to {x^{-\xi} \over \Gamma(1-\xi)},
\ene
where we have dropped possible $\theta$ and $\delta$ functions.  From (59) it is easy to find that, with respect to $m^{\beta}$, the
transformed series is given by
\beg
\hat Z_{N}(x)=\sum^{N}_{n=0} (-1)^n {\Gamma(2n+1/2) \over n!
\Gamma(4n/\beta+1/\beta+1)}x^{4n/\beta+1/\beta}.
\ene
Then we find that $\hat Z_{\infty}(x)$ is a divergent series for $\beta>4$  
and becomes a convergent one for $\beta=4$ where the convergence radius is $1/4$.  When $\beta<4$, the series becomes a convergent one for any $x$.  We
have plotted in Fig.4 the series for $\beta\in [1, 3]$ at $N=100$.  Next problem is whether the limit $\lim_{x\to \infty}\hat Z_{\infty}(x)$ exists or not.  From the numerical calculation at $N=200$, we find that the result changes around $\beta\sim 1.32$:   For $\beta\in [1, 1.32]$, $\hat Z_{\infty}(x)$ would not converge in the $x\to
 \infty$ limit.  The convergence of $\lim_{x\to \infty}\hat Z_{\infty}(x)$ seems to be
realized for $\beta\in [1.32, 4]$.  We do not know why around $\beta=1.32$ the convergence property changes.  Now, we observe the following features:  When $\beta$ is small the truncated series becomes effective to large
$x$ but $\hat Z(x)$ does not converge at $x=\infty$.  As $\beta$ increases beyond $\sim 1.32$ but
still within $\beta<4$, the effective range of truncated series decreases
but the saturation to finite $\hat Z(\infty)$ is fast.  Furthermore the oscillation amplitude damps as $\beta$ decreases.   From these points, it is subtle that,
at a fixed order, what value of $\beta$ between $\sim 1.32$ and $4$ gives the best approximation.  We have calculated $\hat Z_{N}(x^{*})$ 
 at $N=99$ or $100$ for $\beta=1.5, 1.7, 1.9, 2.0$ and $2.1$.  The result reads
\begin{eqnarray}
\beta=1.5&\quad&1.8169575 \quad (x=15.037,\hskip 4pt N=100)\nonumber\\
\beta=1.7&\quad&1.81926459 \quad (x=10.336,\hskip 4pt N=100)\nonumber\\
\beta=1.9&\quad&1.81719914639 \quad (x=7.1806,\hskip 4pt N=100)\nonumber\\
\beta=2.0&\quad&1.812804954110934 \quad (x=5.34,\hskip 4pt N=99)\nonumber\\
\beta=2.1&\quad&1.8051679 \quad (x=5.043,\hskip 4pt N=99).
\end{eqnarray}
Thus we find that the best approximation is realized at $\beta=2.0$.  The agreement to the exact value, $1.812804954110954\cdots$ is remarkable for that case (The first 14 decimals are in the agreement).

We note that the ordinary Borel summation method employees $\lambda$ as the
transformation variable.  This means in our context that $\sigma=m^{4}$.   Then the convergence
radius of perturbative $\hat Z(x)$ becomes
finite ($\rho=1/4$) and the choice is not suited for our modified Laplace representation approach where the
integrand is truncated and the upper integration limit is cut off.  

Now we turn to the anharmonic oscillator.  If one performs
Heaviside transformation with respect to $m^{\beta}$, one has
\beg
\hat E_{N}(x)=\sum_{n=0}^{N}{A_{n} \over
\Gamma(3n/\beta-1/\beta+1)}x^{(3n-1)/\beta}.
\ene
The n-th coefficient behaves for large $n$ as
\beg
{A_{n} \over \Gamma(3n/\beta-1/\beta+1)}\sim (-1)^{n+1}\biggl({3n \over
e}\biggl)^{(1-3/\beta)n}\beta^{(3n-1)/\beta}n^{-1/2+1/\beta}.
\ene
Hence the series is divergent for $\beta>3$ and $\rho$ is infinite for  $\beta<3$. 
For $\beta=3$ the series is a convergent one, but $\rho=1/3$.
  We have figured out $\hat E_{249}(x)$ by varying $\beta$ from $1$ to $3$. 
The result is shown in
Fig.5.  The convergence property of $\hat E_{\infty}(x)$ in the $x\to \infty$ limit seems to change around $\beta\sim 1.15$.  As in the previous case, we examined what value around $\beta=2$ gives the best approximation of $E(0)$ at $N=248$ or $249$.  For $\beta=1.7,1.8,1.9,2.0,2.1$ we obtained
\begin{eqnarray}
\beta=1.7&\quad&0.6655 \quad (x=6.53\sim 6.57,\hskip 4pt N=248)\nonumber\\
\beta=1.8&\quad&0.665514724 \quad (x=5.1257\sim 5.1260,\hskip 4pt N=248)\nonumber\\
\beta=1.9&\quad&0.666234824085 \quad (x=4.0185,\hskip 4pt N=248)\nonumber\\
\beta=2.0&\quad&0.667986259143255939 \quad (x=2.98685,\hskip 4pt N=248)\nonumber\\
\beta=2.1&\quad&0.67107970759 \quad (x=2.5012,\hskip 4pt N=249).
\end{eqnarray}
From the above sample, we find that the choice $\beta=2$ gives the particular good approximation.  
The ordinary Borel choice of variable, $\sigma=m^{3}$, also does not work well in our approach. 
This is because in that choice the convergence radius of perturbative $\hat E(x)$ becomes finite ($\rho=1/3$). 

It is interesting to note that why the choice, $\beta=2$, gives the best approximation in both cases.  Although we have not resolved it yet,    we suspect that the resolution would have related to the fact that $Z(m)$ and $E(m)$ has expansion in the square of the mass, $c_{0}+c_{1}m^2+c_{2}(m^2)^2+\cdots$.  Any finite sum, $\sum_{i=0}^{n}c_{i}(m^2)^{i}$, vanishes, if the integration variable is chosen as $\sigma=m^2$ because from (59),
\beg
\sigma^i\to 0\quad (i=1,2,\cdots).
\ene
Hence, there is no powers of $1/x$ in the corresponding Heaviside function at large $x$.  Note that when the power $\xi$ is fractional the transform of $\sigma^{\xi}$ survives to give $x^{-\xi}$.  
  Thus, as well as $\hat Z(x)$ which is analytically solved as (58), $\hat E(x)$ would be suppressed exponentially only at $\beta=2$ and the smallness of the difference $|\hat E(x)-\hat E(\infty)|$ would be thus explained. 

To conclude this paper, we have demonstrated how truncated perturbation
series can be utilized to approximate the non-perturbative quantities via
modified Laplace representation.  The non-Gaussian integral is precisely
calculated from the truncated perturbation series.  And the approximant for
the strong coupling limit is found to converge in the $N\rightarrow \infty$
limit.  The ground state energy of the anharmonic oscillator is also
approximated successfully over the entire parameter space.  Large order
calculation strongly suggests that our approach works well to all orders.
  In both examples, we find that the perturbative knowledge serves us sufficient information for recovering the small mass or strong coupling expansion.

 In field theories, however, 
perturbation series has more complex structure in general.  For example, the perturbation series for the dressed mass, vacuum energy density or condensates involves the mass logarithm, $\log m$.  When 
the mass-logs exist, it is no longer expected that the convergence
radius is enlarged by Heaviside transformation with respect to the mass, although some qualitative
improvement is found in the literature (first paper in ref.6).  This is the
subject of our future investigation.    

\newpage
\Large
\noindent {\bf  Appendix: Linear $\delta$ expansion in a particular limit and the Heaviside transformation}
\vskip 5pt
\large
\baselineskip 24pt
In this Appendix we 
show that, in a particular limit, the linear $\delta$ expansion leads to the Heaviside transformation.     

In linear $\delta$ expansion, the Lagrangian of anharmonic oscillator is
written by introducing
auxiliary or variational mass $\Omega$ as,
\beg
L={1 \over 2}\biggl({dq \over dt}\biggl)^{2}-{1 \over 2}\Omega^{2}q^{2}-{1
\over
2}(m^{2}-\Omega^{2})q^{2}-\lambda q^{4},
\ene
and the free and interaction parts are defined as
\begin{eqnarray}
L_{free}&=&{1 \over 2}\biggl({dq \over dt}\biggl)^{2}-{1 \over
2}\Omega^{2}q^{2},\nonumber\\
L_{int}&=&{1 \over 2}\delta\Omega^{2}q^{2}-\lambda q^{4},
\end{eqnarray}
where
\beg
\delta=1-m^{2}/\Omega^{2}.
\ene
Let us concentrate on the $m=0$ case hereafter.  Then one should set $\delta=1$,
which represents the masslessness of the starting Lagrangian, at the end of the calculation.
	
First we point out that perturbation expansion in $L_{int}$ is generated
from the ordinary perturbative result with mass $\Omega$ by shifting
$\Omega^{2}\rightarrow \Omega^{2}(1-\delta)$ and then expanding in powers of
$\delta$ to relevant orders.  Let the number of $\delta\Omega^{2}$ vertex and $\lambda$ coupling in a given Feynman diagram be $n_{\Omega}$ and $n_{\lambda}$ respectively.  Then at order $n$ in $L_{int}$, any contributing diagram should obey $n_{\Omega}+n_{\lambda}\le n$.  To proceed to large orders keeping the analytical manipulation straightforward, it is however convenient to modify the  expansion scheme.  The new expansion is defined such that, at order $n$, contributing diagram should obey $n_{\lambda}\le n$ {\it and} $n_{\Omega}\le n$.  Even a diagram has $n$ vertices for $\lambda$ coupling and $n$ mass insertions, it should be included to the $n$-th order expansion.  In this expansion scheme, any contributing diagrams are obtained such that, given a Feynman diagram with no $\delta\Omega^{2}$ vertex, one should incorporate it by shifting $\delta$ and then expanding in $\delta$ up to just the order of perturbative expansion.  This procedure is formally carried out as we can see below.
 
Consider a given Feynman amplitude with no $\delta \Omega^2$ vertex, $f(\Omega^{2})$.   First we shift
$\Omega^{2}\rightarrow \Omega^{2}(1-\delta)$ and expand the result in
$\delta$.   We then have
\beg
f(\Omega^{2}-\Omega^{2}\delta)=\sum^{\infty}_{k=0}f^{(k)}(\Omega^{2})
{(-\Omega^{2}\delta)^{k} \over k!}=\lim_{N\rightarrow
\infty}\sum^{N}_{k=0}{(-\delta\Omega^{2})^{k} \over k!}
\Bigl({\partial \over \partial \Omega^{2}}\Bigl)^{k}f(\Omega^{2}).
\ene
Setting $\delta=1$ we have
\beg
f(0)=\lim_{N\rightarrow \infty}\sum^{N}_{k=0}{(-\Omega^{2})^k \over k!}
\Bigl({\partial \over \partial \Omega^{2}}\Bigl)^{k}f(\Omega^{2}).
\ene
By using
\beg
{\partial \over \partial
p}{1 \over p}={1 \over p}{\partial \over \partial p}-{1 \over p^{2}},
\ene
we find
\beg
\sum^{N}_{k=0}{(-p)^{k} \over
k!}\Bigl({\partial \over \partial p}\Bigl)^{k}=
p{(-p)^{N} \over N!}\Bigl({\partial \over \partial p}\Bigl)^{N}{1 \over p}
\stackrel{\rm def}{=} {\cal D}_{N}(p).
\ene
Thus we arrive at
\beg
f(0)=\lim_{N\rightarrow \infty}{\cal
D}_{N}(\Omega^{2})f(\Omega^{2}).
\ene
This is, however, just a formal result and we address explicit example
and ask what comes out.

The operation of ${\cal D}_{N}(\Omega^2)$ on $(\Omega^2)^{\xi}$ gives
\beg
{\cal D}_{N}(\Omega^2)(\Omega^2)^{\xi}={(N-\xi)\cdots(2-\xi)(1-\xi) \over N!}(\Omega^{2})^{\xi}.
\ene
When $N$ is large enough, the right hand side approaches to $(\Omega^2/N)^{\xi}/\Gamma(1-\xi)$ and the result of operation of ${\cal
D}_{N}(\Omega^{2})$ on $E_{N}(\Omega)$
reads, 
\beg
{\cal D}_{N}(\Omega^{2})E_{N}(\Omega)\sim\sum^{N}_{n=0}{A_{n} \over
\Gamma(3n/2+1/2)}\Bigl({N \over \Omega^2}\Bigl)^{3n/2-1/2}\quad (N\gg 1).
\ene
Thus when $N/\Omega^2$ is changed to $x$ in (75) the result agrees with (43). 
However note that, in accord with $N$, we must let $\Omega$ large enough to
stop the divergence of
RHS of (75) (Note that at $N=249$, the largest stationary point is given at $N/\Omega^{2}\sim 3$).  To give (75)
some meaning, it is thus necessary to let $N$ and $\Omega^{2}$
simultaneousely large with $N/\Omega^{2}$ kept finite.  And accordingly, $\Omega^{2}$ is no longer a free finite parameter.  If we adopt this recipe, the equality of (73) breaks down and we should take the righthand side of (73) as a new function $\hat f$ of variable $x=N/\Omega^2$.  We can expect the equality relation only in the limit, $x\to \infty$, such that $\lim_{x\to \infty}\hat f(x)=f(0)$ (This limit corresponds to $N\to \infty$ with $\Omega^2$ fixed).

Now we show that ${\cal D}_{N}$
induces Heaviside function in the limit that
$N, \Omega\rightarrow \infty$ with $\Omega^{2}/N$ fixed.  Let us start with
the Laplace transform,
\beg
f(\Omega)=\Omega^{2}\int^{\infty}_{-\infty}dt \exp(-\Omega^{2} t)\hat f(t).
\ene
By operating ${\cal D}_{N}$ we have
\beg
{\cal D}_{N}f(\Omega)=\int^{\infty}_{-\infty}dt {(\Omega^{2})^{N+1}t^{N}
\over
N!}\exp(-\Omega^{2}
t)\hat f(t)\stackrel{\rm def}{=} \int^{\infty}_{-\infty}dt \Delta_{N,\Omega^{2}}(t)\hat
f(t).
\ene
We find that the function $\Delta_{N,\Omega^{2}}(t)$ approaches to the Dirac
$\delta$ function 
in the limit, $N,\Omega\rightarrow \infty$ with the ratio
$N/\Omega^{2}$ fixed,
\beg
\Delta_{N,\Omega^{2}}(t)\rightarrow \delta(t-N/\Omega^{2}).
\ene
Thus we find that 
\beg
\lim_{N,\Omega^{2}\rightarrow \infty}{\cal D}_{N}f(\Omega)=\hat
f(N/\Omega^{2}).
\ene
This result states that ${\cal D}_{N}$ gives the kernel of Laplace transform in the limit $N,\Omega\rightarrow \infty$ with the ratio
$N/\Omega^{2}$ fixed.

We remark that although we have shown that, in our modified expansion scheme, the $\delta$ expansion can induce the Heaviside transform in large orders, this may not lead that the conventional expansion in $L_{int}$ also gives the same function $\hat f$ in the large order limit.  This is because the difference of the included diagrams between the two expansions increases as the order increases.

\newpage
\begin{center}
{\bf References}
\end{center}
\begin{description}
\item [{1}] C. M. Bender and T. T. Wu, Phys. Rev. 184 (1969) 1231; Phys.
Rev.
D7 (1973) 1620.
\item [{2}] For an earler review see F. T. Hioe, D. McMillen and E. W.
Montroll, Phys. Rep. 43C (1978) 305.

\item [{3}] See for earlier references, W. E. Caswell, Ann. Phys. 123 (1979)
153; J. killingbeck, J. Phys. A14 (1981) 1005; R.Seznec and J. Zinn-Justin,
J.
Math. Phys. 20 (1979) 1398; P. M. Stevenson, Phys. Rev. D23 (1981) 2916;
 L. N. Chang and N. P. Chang, Phys. Rev. D29 (1984) 312; A. Okopinska, Phys. Rev. D35 (1987); A. Duncan and M. Moshe, Phys. Lett. B215
(1988) 352; A. Neveu, Nucl. Phys. B(proc. Suppl.) 18B (1991) 242; H. Yamada, Mod. Phys. Lett. A6 (1991) 3405; G. Klimenko, Z. Phys. C50 (1991) 477.  Further
references are found in ref.5.
\item [{4}] W. Janke and H. Kleinert, Phys. Rev. Lett. 75 (1995) 2787.
\item [{5}] B. Bellet, P. Garcia and A. Neveu, Int. J. Mod. Phys. A11 (1996) 5587.
\item[{6}] A. Duncan and H. F. Jones, Phys. Rev. D47 (1993) 2560; R. Guida, K.Konishi and H. Suzuki, Ann. Phys. 241 (1995) 152; Ann. Phys. 249 (1996)109; C. Arvanitis, H. F. Jones and C. S. Parker, Phys. Rev. D52 (1995) 3704.
\item [{7}] H. Yamada, Mod. Phys. Lett. A11 (1996) 1001;  Mod. Phys. Lett. A11 (1996) 2793.
\item [{8}] S. Moriguchi et al., Suugaku Koushiki II, Iwanami Shoten (in
Japanese).
\item [{9}] S. Graffi, V. Grecchi and B. Simon, Phys. Lett. 32B (1970) 631.
\item [{10}] H. Yamada, Mod. Phys. Lett. A9 (1994) 1195; Int. J. Mod. Phys.
A9 (1994) 5651.
\item [{11}] R. Balsa, M. Plo,
J. G. Esteve and A. F. Pacheco, Phys. Rev. D28 (1983) 1945.
\item [{12}] F. Vinette and J. Cizek, J. Math. Phys. 32 (1991) 3392. 
\end{description} 
\newpage  

\begin{center}
{\bf Table Captions}
\end{center}
\begin{description}
\item [{Table 1}] Numerical computation of $Z_{N}(x^{*})$ to 15-th orders. 
\item [{Table 2}] Numerical computation of $Z_{15}(m^{2},x^{*})$ for various values of $m^{2}$.  At this order $x^{*2}=5.0438870$.
\item [{Table 3}] Low order approximation of first several strong coupling
coefficients of the ground state energy.
\item [{Table 4}] Numerical result of $E_{249}(m^{2},x^{*})$ for various $m^{2}$.
\end{description}   

\begin{center}
{\bf Figure Captions}
\end{center}
\begin{description}
\item [{Fig. 1}] The perturbative Heaviside functions $\hat Z_{N}(x)$ for
$N=1,4,7$. 
\item [{Fig. 2}] The ratio, $E_{N}(m,x^{*})/E_{exact}(m)$ is plotted at
$N=1$ and $5$.
\item [{Fig. 3}] The function $\hat E_{249}(x)$ is shown.  The sharp rise
around $x\sim 3.2$ represents the break down of the perturbative truncation.
\item [{Fig. 4}] The function $\hat Z_{100}(x)$ is shown for various choice of $\beta$.  The amplitude of oscillation of $\hat Z_{100}(x)$  decreases as $\beta$ decreases from $1.0$ to $2.0$.  The effective range of the series is larger for smaller $\beta$ but the saturation to $\hat Z(\infty)$ is slower.  When $\beta$ is close to $2.0$, the oscillation is weak and the saturation to $\hat Z(\infty)$ is very fast.
\item [{Fig. 5}] The function $\hat E_{249}(x)$ is shown for various
$\beta$.  At $\beta=1$, $\hat E_{N}(x)\to 0$ as $x\to 0$.  When $1<\beta\le 3$, $\hat E_{N}(x)$ diverges in the $x\to 0$ limit.  The oscillation amplitude of $\hat E_{249}(x)$ damps as $\beta$ decreases.  At $\beta=2.0$, the oscillation is quite weak and the saturation to $\hat E(\infty)$ is very fast.
\end{description}   

\begin{center}
\begin{tabular}{lcc} \hline {\it N} & {$\hat Z_{N}(x^{*})$} & {$x^{*2}$}\\
\hline
1 & 1.6 & 1 \\
3 & 1.7313594 & 1.5960716 
\\ 
5 & 1.7765256 & 2.1806071 \\ 
7 & 1.7955618 & 2.7590027 
\\ 
9 & 1.8043006 & 3.3335514 \\
11 & 1.8085078 & 3.9054517
\\ 
13 & 1.8105959 & 4.4754119
\\ 
15 & 1.8116546 & 5.0438870
  
\\  
exact & 1.8128049 &
 
 \\ 
\hline
\end{tabular}
\end{center}
\centerline{\Large Table 1}
\vskip 1cm
\begin{center}
\begin{tabular}{lcc} \hline {$ m^2$} & {$Z_{15}(m^{2},x^{*})$} & {$exact$}\\
\hline
0.01 & 1.80557702921362946 & 1.806700454307384679\\
0.1 & 1.75281762908207768 & 1.753725831772014832 
\\ 
1 & 1.36831695165151724 & 1.368426855735508774 \\ 
3 & 0.96173724333279584 & 0.961738333157472108 
\\ 
6 & 0.710038679143677110 & 0.710038680405132855\\
10 & 0.556465718382570615 & 0.556465718382772753 
\\ 
100 & 0.177232097497741759 & 0.177232097497741761  
\\  
\hline
\end{tabular}
\end{center}
\centerline{\Large Table 2}
\vskip 1cm
\begin{center}
\begin{tabular}{lccc} \hline {\it N} & {$\hat E(x^{*})$} & {$\alpha_{1}$} &
{$\alpha_{2}$}\\ \hline
1 & 0.738558766382022 & 0.121215344755496 & -0.004420970641441 \\
3 & 0.686283726385561 & 0.134984882799344 & -0.006437328047158
\\ 
5 & 0.674564660427775 & 0.139821127686036 & -0.007465973798930  \\ 
7 & 0.670682699394559 & 0.141853009994811 & -0.008005281363443 
\\ 
9 & 0.669175108154224 & 0.142780937393257 & -0.008293166786158  \\ 
exact & 0.667986259155... &
0.143668783380...
 & -0.008627565680...  \\ 
\hline
\end{tabular}
\vskip 5mm
\begin{tabular}{lccc} \hline {\it N} & {$\alpha_{3}$} & {$\alpha_{4}$} &
{$x^{*}$} \\ \hline
1 &
 0.0002176764415436  & -0.000010207418675 & 0.328248340614232 \\
3 &
 0.0004280233911678 &  -0.000027447789756 & 0.448360373271548\\ 
5 &
 0.0005773467460109 &  -0.000044015663387 & 0.549152913559036 \\ 
7 & 
 0.0006739106832314 &  -0.000057114935211 & 0.636621775859320\\ 
9 &
 0.0007338756445212 &  -0.000066541377300 & 0.713814879924458\\ 
exact  &
 0.000818208905... & -0.000082429217...   \\ 
\hline
\end{tabular}
\end{center}
\centerline{\Large Table 3}
\vskip 1cm
\begin{center}
\begin{tabular}{lcc} \hline {$m^{2}$} & {$E_{249}(m^{2},x^{*})$} & {exact}
\\
\hline
0.001  & 0.66812991929974827 & 0.66812991931241042 \\
0.01 &   0.66942208503810206 & 0.66942208505040309  
\\ 
0.1 &    0.68226767187380087 & 0.68226767188301217 \\ 
1 &      0.80377065123375873 & 0.80377065123427376 
\\ 
10 &     1.64938954183035211 & 1.64938954183035211\\
100 &    5.00747395574729234 & 5.00747395574729234
\\ 
1000 &   15.8121382178529  & 15.8121382178529
\\ 
\hline
\end{tabular}
\end{center}
\centerline{\Large Table 4}
\end{document}